\documentclass[twocolumn,prl,showpacs,amsfonts,amsmath,assymb,eufrak,superscriptaddress]{revtex4-1}
\usepackage{amsmath,amsfonts,amssymb,mathtools,upgreek}
\usepackage{epsfig}

\usepackage[dvipdfmx]{hyperref}\hypersetup{hidelinks}

\def\eq#1\en{\begin{equation}#1\end{equation}} 
\def\eqa#1\ena{\begin{align}#1\end{align}}
\def\eqg#1\eng{\begin{gather}#1\end{gather}}
\newcommand{\lb}[1]{\label{e:#1}}
\newcommand{\rlb}[1]{\eqref{e:#1}} 
\newcommand{\nl}{\notag\\}

\newcommand{\snorm}[1]{\Vert#1\Vert}

\newcommand{\sbkt}[1]{\langle#1\rangle}

\newcommand{\sumtwo}[2]%
{\mathop{\sum_{#1}}_{#2}}
\newcommand{\sumthree}[3]%
{\mathop{\mathop{\sum_{#1}}_{#2}}_{#3}}
\newcommand{\sumfour}[4]%
{\mathop{\mathop{\mathop{\sum_{#1}}_{#2}}_{#3}}_{#4}} 
\newcommand{\prodtwo}[2]%
{\mathop{\prod_{#1}}_{#2}}
\newcommand{\mintwo}[2]%
{\mathop{\min_{#1}}_{#2}}
\newcommand{\maxtwo}[2]%
{\mathop{\max_{#1}}_{#2}}
\newcommand{\maxthree}[3]%
{\mathop{\mathop{\max_{#1}}_{#2}}_{#3}}
\newcommand{\limtwo}[2]%
{\mathop{\lim_{#1}}_{#2}}
\newcommand{\suptwo}[2]%
{\mathop{\sup_{#1}}_{#2}}
\newcommand{\supthree}[3]%
{\mathop{\mathop{\sup_{#1}}_{#2}}_{#3}}
\newcommand{\supfour}[4]%
{\mathop{\mathop{\mathop{\sup_{#1}}_{#2}}_{#3}}_{#4}} 
\newcommand{\inftwo}[2]%
{\mathop{\inf_{#1}}_{#2}}
\newcommand{\infthree}[3]%
{\mathop{\mathop{\inf_{#1}}_{#2}}_{#3}}
\newcommand{\inffour}[4]%
{\mathop{\mathop{\mathop{\inf_{#1}}_{#2}}_{#3}}_{#4}} 
\newcommand\calA{{\cal A}}

\newcommand\calK{{\cal K}}

\newcommand{\bsk}{\boldsymbol{k}}

\newcommand{\up}{\uparrow}

\newcommand{\La}{\Lambda}

\newcommand{\bra}[1]{\langle#1|}
\newcommand{\ket}[1]{|#1\rangle}

\newcommand{\la}{\lambda}

\newcommand{\hA}{\hat{A}}
\newcommand{\hP}{\hat{P}}
\newcommand{\hH}{\hat{H}}
\newcommand{\hN}{\hat{N}}
\newcommand{\hn}{\hat{n}}

\newcommand{\had}{\hat{a}^\dagger}
\newcommand{\hrho}{\hat{\rho}}

\newcommand{\kPz}{\ket{\Phi(0)}}

\newcommand{\Pneq}{\hP_{\rm neq}}

\newcommand{\hbd}{\hat{b}^\dagger}
\newcommand{\hb}{\hat{b}}
\newcommand{\hcd}{\hat{c}^\dagger}
\newcommand{\hc}{\hat{c}}

\newcommand{\vac}{\ket{\Phi_{\rm vac}}}
\newcommand{\vacb}{\bra{\Phi_{\rm vac}}}

\newcommand{\para}[1]{\medskip\par{\em #1}\/.---}

\begin{document}
\title{Macroscopic Irreversibility in Quantum Systems: Free Expansion in a Fermion Chain}

\author{Hal Tasaki}\email[]{hal.tasaki@gakushuin.ac.jp}
\affiliation{Department of Physics, Gakushuin University, Mejiro, Toshima-ku, Tokyo 171-8588, Japan}

\date{\today}

\begin{abstract}
We consider a free fermion chain with uniform nearest-neighbor hopping and let it evolve from an arbitrary initial state with a fixed macroscopic number of particles.
We then prove that, at a sufficiently large and typical time, the measured coarse-grained density distribution is almost uniform with (quantum mechanical) probability extremely close to one.
This establishes the emergence of irreversible behavior, i.e., a ballistic diffusion, in a system governed by quantum mechanical unitary time evolution.
It is conceptually important that irreversibility from any initial state is proved here without introducing any randomness to the initial state or the Hamiltonian, while the known examples, both classical and quantum, rely on certain randomness or apply to limited classes of initial states.
The essential new ingredient in the proof is the large deviation bound for every energy eigenstate, which is reminiscent of the strong ETH (energy eigenstate thermalization hypothesis).

\par\noindent
{\em There is a 20-minute video that explains the main results of the present work:} 
 \\\url
 {https://youtu.be/HQzzCxok18A}

\end{abstract}

\maketitle
The emergence of macroscopic irreversibility in physical systems governed by reversible time-evolution laws is a fundamental and fascinating topic in modern physics \cite{Lebwitz1993}.
It is now understood that the most important factor for irreversible behavior is a large degree of freedom.
In fact, irreversibility can be observed even in ideal gases, provided that the number of particles is large \cite{BievreParris2017,RigolMuramatsuOlshanii2006,RigolFitzpatrick2011,Pandeyetal,GluzaEisertFarrelly2019,ShiraishiTasaki2023,LydzbaZhangRigolVidmar2021,LydzbaMierzejewskiRigolVidmar2023}.

As a simple but illuminating example, consider a system of $N$ free classical particles, where $N\gg1$, on the interval $[0,L]$ with periodic boundary conditions.
All particles are initially at the origin, and the velocity of the $j$-th particle is $v_j$.
The position of the $j$-th particle at time $t$ is then given by $x_j(t)=tv_j\ {\rm mod}\,L$.
Suppose that the initial velocities are drawn randomly, independently, and uniformly from the interval $[-v_0,v_0]$.
Then, for $t$ such that $tv_0\gg L$, the positions $x_j(t)$ of the particles are almost uniformly distributed in the interval $[0,L]$.
This is clearly seen by introducing the coarse-grained density distribution by decomposing $[0,L]$ into $m=O(1)$ identical intervals and examining the particle densities in the $m$ intervals.
One observes an evolution from a concentrated distribution at $t=0$ to an almost uniform distribution at large $t$.
We conclude that this classical deterministic system exhibits a macroscopic irreversible behavior.
See \cite{BievreParris2017} and references therein for related results.

It should be noted that the above statement is intrinsically probabilistic in the sense that the irreversible behavior is shown to take place (for each fixed large $t$) for the majority of random initial velocities.
One may fail to observe irreversibility if the velocities are chosen according to a deterministic rule; an extreme example is the case where all $v_j$ are identical.
More generally, one needs a random initial state to observe macroscopic irreversibility in most deterministic classical mechanical systems, including those exhibiting (quasi)periodic or chaotic behavior \cite{Lebwitz1993,BievreParris2017,GuckenheimerHolmes,Gutzwiller,EckmannRuelle}.

Interestingly, the situation can be essentially different for quantum systems.
In the present paper, we prove that a free fermion chain with a macroscopic number of particles exhibits irreversible expansion.
More precisely, we show that the system, starting from any initial state and evolving under the unitary time evolution, is found in a state with uniform density at a sufficiently large and typical time.
This rigorously establishes that a non-random quantum many-body system can exhibit macroscopic irreversibility from an arbitrary initial state.
This is in contrast to previously known rigorous examples of macroscopic irreversibility, both in classical and quantum systems, which rely either on randomness in the system or the random (or restricted) choice of initial states \cite{Lebwitz1993,BievreParris2017,GluzaEisertFarrelly2019,ShiraishiTasaki2023}.

Our proof, although rather short, is based on an accumulation of ideas and methods developed over the decades to understand thermalization in isolated macroscopic quantum systems, especially those related to ETH (energy eigenstate thermalization hypothesis). 
See, e.g., \cite{vonNeumann,Deutsch1991,Srednicki1994,Tasaki1998,GLTZ,GLMTZ09b,RigolSrednicki2012,Reimann3,DAlessioKafriPolkovnikovRigol2016,Tasaki2016}.
The most important strategy is ETH in the large-deviation form formulated by us in \cite{Tasaki2016}.
Essential ingredients specific to the free fermion chain are the absence of degeneracy in the many-body spectrum (Lemma~1) proved in our previous works \cite{Tasaki2010,Tasaki2016,ShiraishiTasaki2023}, and the large deviation bound reminiscent of the strong ETH (Lemma~4) proved in this paper.

Although we here present a single specific example, we believe our findings will shed light on the problems of irreversiblity and thermalization in a more general class of quantum many-body systems.
See Discussion.

\para{The model and the absence of degeneracy}%
Consider a system of $N$ spinless fermions on the chain $\La=\{1,2,\ldots,L\}$ with periodic boundary conditions.
We fix $L$ to be a large prime number and fix large $N$ such that $N<L$.
The density is $\rho_0=N/L$.

Let $x,y,\ldots\in\La$ denote lattice sites and $\hc_x$, $\hcd_x$, and $\hn_x=\hcd_x\hc_x$ denote the annihilation, creation, and number operators, respectively, of the fermion at site $x\in\La$.
We take the standard free fermion Hamiltonian
\eq
\hH=\sum_{x=1}^L\bigl\{e^{i\theta}\,\hcd_x\hc_{x+1}+e^{-i\theta}\,\hcd_{x+1}\hc_x\bigr\},
\lb{H}
\en
where we introduced an artificial flux $\theta\in[0,2\pi)$ to avoid degeneracy.  See Lemma~1 below.

The Hamiltonian $\hH$ is readily diagonalized in terms of the plane wave states.
Setting the $k$-space as
$\calK=\{(2\pi/L)\nu\,|\,\nu=0,\ldots,L-1\}$,
we define the creation operator
$\had_k=L^{-1/2}\sum_{x=1}^Le^{ikx}\,\hcd_x$
for $k\in\calK$.
Let $\bsk=(k_1,\ldots,k_N)$ denote a collection of $N$ elements in $\calK$ such that $k_j<k_{j+1}$.
Then
\eq
\ket{\Psi_{\bsk}}=\had_{k_1}\had_{k_2}\cdots\had_{k_N}\vac,
\lb{Psik}
\en
is an eigenstate of $\hH$ with the eigenvalue
\eq
E_{\bsk}=2\sum_{j=1}^N\cos(k_j+\theta).
\lb{Ek}
\en
These are the only energy eigenstates and eigenvalues.

Then we have the following essential fact proved in our previous works \cite{Tasaki2010,Tasaki2016,ShiraishiTasaki2023}.

\para{Lemma 1}
Let $L$ be an odd prime.
For any $N$, the energy eigenvalues \rlb{Ek} are non-degenerate except for a finite number of $\theta$.
In particular the spectrum is free from degeneracy if $\theta\ne0$ satisfies $|\theta|\le(4N+2L)^{-(L-1)/2}$.

\medskip
From now on, we shall only consider the case in which the absence of degeneracy is guaranteed by the above lemma.
It suffices to set, e.g.,  $\theta=(4N+2L)^{-(L-1)/2}$.

\para{Main results}
For a fixed integer $m\ge2$ of order one, let $I_1,\ldots,I_m\subset\La$ be non-overlapping intervals containing $\ell=\lfloor\frac{L}{m}\rfloor$ sites.
Note that $\bigcup_{j=1}^mI_j$ almost covers the whole chain $\La$.
Let $\hN_j=\sum_{x\in I_j}\hn_x$  be the number of particles in $I_j$.
We suppose that a macroscopic observer makes a simultaneous measurement of the densities $\hrho_j=\hN_j/\ell$ for $j=1,\ldots,m$ to determine the coarse-grained density distribution.
The observer concludes that the system is in equilibrium if the measured densities $\rho_j$ satisfy $|\rho_j-\rho_0|\le\rho_0\delta$ for all $j$, where $\delta\in(0,\frac{1}{2}]$ is an arbitrary fixed (small) relative density.
This motivates us to define the nonequilibrium projection $\Pneq$ as the projection operator onto the subspace characterized by $|\hrho_j-\rho_0|\ge\rho_0\delta$ for some $j$.
Then $\bra{\Phi}\Pneq\ket{\Phi}$ is the probability that the observer finds the state $\ket{\Phi}$ not in equilibrium.

We stress that our criterion for equilibrium is based solely on the macroscopic measurement.
This is a natural strategy for studying the emergence of macroscopic physics from microscopic quantum physics.

We first present a preliminary ``ergodic version" of our result.
Here, $\ket{\Phi(t)}=e^{-i\hH t}\kPz$ denotes a general solution of the Schr\"odinger equation.
\para{Theorem 2}
For any initial state $\kPz$ with $N$ particles, we have
\eq
\lim_{T\up\infty}\frac{1}{T}\int_0^T\!\!\!dt\,\bra{\Phi(t)}\Pneq\ket{\Phi(t)}\le e^{-\frac{\delta^2}{3(m-1)}N}.
\lb{main1}
\en
\medskip

Note that we have $\delta^2N\gg1$ in the natural setting where $\delta$ is small (but $N$ independent) and $N$ is macroscopically large.
Then we see that the right-hand side of \rlb{main1} is negligibly small, and hence, the system is essentially in equilibrium in the long-time average.

We note that Theorem~2 most clearly highlights the essential difference between irreversibility in classical and quantum systems.
Recall that the ergodicity in a classical deterministic system deals with a property that is valid with probability one with respect to a specified probability distribution of the initial state \cite{EckmannRuelle}.
In contrast, our ergodicity statement \rlb{main1} is valid for every initial state.

By a standard argument, we can convert Theorem~2 into the following statement relevant to the instantaneous measurement of the coarse-grained density distribution.

\para{Theorem 3}
For any initial state $\kPz$ with $N$ particles and any sufficiently large $T>0$, there exists a subset $\calA\subset[0,T]$ with
${l(\calA)}/{T}\le e^{-\frac{\delta^2}{8(m-1)}N}$
(where $l(\calA)$ is the total length or the Lebesgue measure of $\calA$) such that we have
\eq
\bra{\Phi(t)}\Pneq\ket{\Phi(t)}\le e^{-\frac{\delta^2}{8(m-1)}N},
\lb{main2}
\en
for any $t\in[0,T]\backslash\calA$.

\medskip
Here, how large $T$ should be depends on $\kPz$.

Again, assume $\delta^2N\gg1$.
Then \rlb{main2} shows that the coarse-grained density distribution obtained from the state $\ket{\Phi(t)}$ is uniform (within the precision $\delta$) with probability very close to one (more precisely, not less than $1-e^{-\frac{\delta^2}{8(m-1)}N}$), provided that $t\in[0,T]\backslash\calA$.
We also see that almost all $t$ in $[0,T]$ belongs to $[0,T]\backslash\calA$ since $l(\calA)/T$ is also exponentially small.
We can say that $\calA$ is the set of atypical moments.

Informally speaking, Theorem 3 establishes that, for a sufficiently large and typical time $t$, the measured coarse-grained density distribution in the time-evolved state $\ket{\Phi(t)}$ is almost certainly uniform.
It is essential here that we are dealing with the result of a single quantum mechanical (simultaneous) measurement rather than a quantum mechanical average (which is obtained through repeated measurements in an ensemble of states).
We conclude that the free fermion chain exhibits irreversible expansion for any initial state $\kPz$ in which $\bra{\Phi(0)}\hrho_j\kPz$ differs from $\rho_0$ for some $j$.

\para{ETH-like bound and proof of Theorem 2}
The following Lemma is a technical key of the present paper.
\para{Lemma 4}
For any energy eigenstate \rlb{Psik} with $N$ particles, we have
\eq
\bra{\Psi_{\bsk}}\Pneq\ket{\Psi_{\bsk}}\le e^{-\frac{\delta^2}{3(m-1)}N}.
\lb{ETH}
\en

\medskip
The bound \rlb{ETH} states that every energy eigenstate is in equilibrium.
It is reminiscent of the strong ETH, especially in its large deviation form discussed in \cite{Tasaki2016}.

Given Lemmas 1 and 4, Theorem 2 is readily proved as follows.
By expanding the normalized initial state in terms of energy eigenstates as $\kPz=\sum_{\bsk}\alpha_{\bsk}\ket{\Psi_{\bsk}}$, the expectation value in the integrand of \rlb{main1} is written as
\eq
\bra{\Phi(t)}\Pneq\ket{\Phi(t)}=\sum_{\bsk,\bsk'}e^{i(E_{\bsk}-E_{\bsk'})t}\alpha_{\bsk}^*\alpha_{\bsk'}\bra{\Psi_{\bsk}}\Pneq\ket{\Psi_{\bsk'}}.
\lb{PPP}
\en
Since Lemma 1 (and our choice of $\theta$) guarantees $E_{\bsk}\ne E_{\bsk'}$ whenever $\bsk\ne\bsk'$, the long-time average of \rlb{PPP} becomes
\eq
\lim_{T\up\infty}\frac{1}{T}\int_0^T\!\!\!dt\bra{\Phi(t)}\Pneq\ket{\Phi(t)}=\sum_{\bsk}|\alpha_{\bsk}|^2\bra{\Psi_{\bsk}}\Pneq\ket{\Psi_{\bsk}}.
\lb{PPP3}
\en
We then get the desired \rlb{main1} from \rlb{ETH}.

Let us prove Lemma 4.
We shall show below an essential inequality
\eq
\bra{\Psi_{\bsk}}e^{\la\hN_j}\ket{\Psi_{\bsk}}\le\{\mu\,e^\la+(1-\mu)\}^N,
\lb{elaN}
\en
for any $j=1,\ldots,m$ and $\la\in(0,1]$, where $\mu=\ell/L\simeq1/m$.
Given \rlb{elaN}, the rest of the estimate is only technical.
Note first that
\eqa
\bra{\Psi_{\bsk}}\hP[\hrho_j-\rho_0\ge\rho_0\delta]\ket{\Psi_{\bsk}}
&\le
\bra{\Psi_{\bsk}}e^{\la(\hN_j-\mu N-\mu N\delta)}\ket{\Psi_{\bsk}}
\nl&\le\bigl\{g(\la,\mu)\,e^{-\la\mu\delta}\bigr\}^N
\lb{LD1}
\ena
with $g(\la,\mu)=\{\mu\,e^\la+(1-\mu)\}e^{-\la\mu}$.
(We denote by $\hP[\cdots]$ the projection onto the specified subspace.)
A straightforward expansion shows $g( \la,\mu)=\sum_{n=0}^\infty\alpha_n\,\la^n$ with $\alpha_n=\{\mu(1-\mu)^n+(-\mu)^n(1-\mu)\}/n!$.
Noting $\alpha_0=1$, $\alpha_1=0$, $\alpha_2=\mu(1-\mu)/2$, and $\alpha_n\le\mu(1-\mu)/n!$ for $n\ge3$, we see
\eqa
g(\la,\mu)&\le1+\frac{\mu(1-\mu)}{2}\la^2+\mu(1-\mu)\la^2\sum_{n=3}^\infty\frac{1}{n!}
\nl
&=1+(e-2)\mu(1-\mu)\la^2\le e^{(e-2)\mu(1-\mu)\la^2}
\lb{gboud}
\ena
Let us set $\la_0=\mu\delta/\{2(e-2)\mu(1-\mu)\}$, which is in $(0,1]$ for $\delta\in(0,\frac{1}{2}]$.
Then we see from \rlb{gboud} that
\eq
g(\la_0,\mu)\,e^{-\la_0\mu\delta}\le e^{-\frac{\mu\delta^2}{4(e-2)(1-\mu)}}.
\en
Recalling that \rlb{LD1} is valid for any $\la\in(0,1]$, we have shown that
\eq
\bra{\Psi_{\bsk}}\hP[\hrho_j-\rho_0\ge\rho_0\delta]\ket{\Psi_{\bsk}}\le e^{-\frac{\mu\delta^2}{4(e-2)(1-\mu)}N}.
\lb{LD2}
\en
By repeating the same analysis for $\hN'_j=N-\hN_j$, we find
\eq
\bra{\Psi_{\bsk}}\hP[\hrho_j-\rho_0\le-\rho_0\delta]\ket{\Psi_{\bsk}}\le e^{-\frac{\mu\delta^2}{4(e-2)(1-\mu)}N}.
\lb{LD3}
\en
Noting that $\Pneq\le\sum_{j=1}^m\hP[|\hrho_j-\rho_0|\ge\rho_0\delta]$,
we get
\eq
\bra{\Psi_{\bsk}}\Pneq\ket{\Psi_{\bsk}}\le2m\,e^{-\frac{\mu\delta^2}{4(e-2)(1-\mu)}N}
\le e^{-\frac{\delta^2}{3(m-1)}N},
\en
where we simplified the bound by noting $4(e-2)\simeq 2.9$ and $\mu\ge\frac{1}{m}-O(\frac{1}{L})$.

It remains to show \rlb{elaN}.
We use an idea similar to that in our previous work \cite{ShiraishiTasaki2023}.
Fix $j$, and note that
\eq
e^{\la\hN_j/2}\ket{\Psi_{\bsk}}=\hbd_{k_1}\hbd_{k_2}\cdots\hbd_{k_N}\vac
\en
with
\eq
\hbd_k=\frac{1}{\sqrt{L}}\Bigl\{e^{\la/2}\sum_{x\in I_j}e^{ikx}\,\hcd_x+\sum_{x\in\La\backslash I_j}e^{ikx}\,\hcd_x\Bigr\}.
\en
We then observe that
\eqa
\bra{\Psi_{\bsk}}e^{\la\hN_j}\ket{\Psi_{\bsk}}&=\vacb\hb_{k_N}\cdots\hb_{k_2}\hb_{k_1}\hbd_{k_1}\hbd_{k_2}\cdots\hbd_{k_N}\vac
\nl
&\le\snorm{\hb_{k_1}\hbd_{k_1}}\vacb\hb_{k_N}\cdots\hb_{k_2}\hbd_{k_2}\cdots\hbd_{k_N}\vac
\nl
&\le\prod_{j=1}^N\snorm{\hb_{k_j}\hbd_{k_j}},
\lb{bbbb}
\ena
where we used the basic property $\bra{\Phi}\hA\ket{\Phi}\le\snorm{\hA}\sbkt{\Phi|\Phi}$ of the operator norm repeatedly.
Noting that $\snorm{\hb_{k}\hbd_{k}}=\{\hbd_k,\hb_k\}=\mu\,e^\la+(1-\mu)$, we get the desired \rlb{elaN}.
(In general, $\{\hbd,\hb\}=\alpha$ and $\hb^2=0$ imply $(\hb\hbd)^2=\alpha\,\hb\hbd$, which means the eigenvalues of $\hb\hbd$ are 0 and $\alpha$.)

\para{Proof of Theorem 3}
The proof is standard \cite{Tasaki2016}, but let us include it for completeness.
From the bound \rlb{main1}, we see that there exists $T_0$ such that
\eq
\frac{1}{T}\int_0^T\!\!\!dt\bra{\Phi(t)}\Pneq\ket{\Phi(t)}\le e^{-\frac{\delta^2}{4(m-1)}N},
\lb{PPP2}
\en
for any $T\ge T_0$.
For given $T\ge T_0$, define
\eq
\calA=\bigl\{t\in[0,T]\,\bigl|\,\bra{\Phi(t)}\Pneq\ket{\Phi(t)}>e^{-\frac{\delta^2}{8(m-1)}N}\bigr\}.
\en
The measure of the set $\calA$ is readily evaluated as
\eqa
&l(\calA)=\int_0^T\!\!\!dt\,\chi\bigl[\bra{\Phi(t)}\Pneq\ket{\Phi(t)}\,e^{\frac{\delta^2}{8(m-1)}N}>1\bigr]
\nl&\le\int_0^T\!\!\!dt\,\bra{\Phi(t)}\Pneq\ket{\Phi(t)}\,e^{\frac{\delta^2}{8(m-1)}N}
\le e^{-\frac{\delta^2}{8(m-1)}N},
\ena
where $\chi[\text{true}]=1$ and $\chi[\text{false}]=0$.
We here used $\chi[x>1]\le x$ and  \rlb{PPP2}.

\para{Discussion}
We proved that a free fermion chain governed by the quantum mechanical unitary time-evolution exhibits an irreversible expansion, as summarized in Theorem~3.
It is of fundamental importance that irreversibility from an arbitrary initial state is established without introducing any randomness in the problem.
Our proof is based on general ideas developed to understand thermalization in isolated quantum systems and also on results specific to free fermion chains.
The essential new ingredient is the strong ETH-like bound stated as Lemma~4.

We admit that our example is fine-tuned; we assume the system size $L$ to be a prime and insert flux $\theta$.
We believe that these unphysical assumptions merely reflect the lack of mathematical techniques. 
The same irreversible behavior should be observed robustly in models without fine-tuning.
In fact, after the earlier version of the present paper appeared in arXiv, it was proved in  \cite{RoosTeufelTumulkaVogel2024} that the same model without flux $\theta$ exhibits similar (but slightly weaker) irreversibility.
Moreover, irreversible behavior in a similar free fermion chain without fine-tuning was numerically observed in  \cite{RigolMuramatsuOlshanii2006}.

We also note that the macroscopic irreversibility observed in our free fermion chain should not be regarded as thermalization in the strict sense, namely, the approach to full thermal equilibrium.
The stationary state obtained in the long-time limit from the initial state $\kPz=\sum_{\bsk}\alpha_{\bsk}\ket{\Psi_{\bsk}}$ is given by the so-called diagonal distribution $\hat{\uprho}_{\rm diag}=\sum_{\bsk}|\alpha_{\bsk}|^2\ket{\Psi_{\bsk}}\bra{\Psi_{\bsk}}$, as can be seen from \rlb{PPP3}.
Note that $\hat{\rho}_{\rm diag}$ may drastically differ from a thermal state.
This is most clearly seen from the fact that the average number of particles with momentum $k\in\calK$ given by
\eq
n_k=\sum_{\bsk}\sum_{j=1}^N\delta(k,k_j)\,|\sbkt{\Psi_{\bsk}|\Phi(t)}|^2,
\en
where the first sum is over $\bsk=(k_1,\ldots,k_N)$, is independent of $t$.
If we chose the initial state $\kPz$ with $n_k$ different from a thermal distribution, the state $\ket{\Phi(t)}$ never settles to thermal equilibrium.

The lack of thermalization in our system reflects the fact that free fermion models are integrable.
It is conjectured that a non-integrable system generally satisfies the strong ETH \cite{DAlessioKafriPolkovnikovRigol2016}, namely, the property corresponding to Lemma~4, not only for the coarse-grained density distribution but for any macroscopic observables.
Then, such a system should exhibit thermalization.

It is, therefore, highly desirable to extend our theory to non-integrable quantum systems, where we expect full-fledged thermalization.
Of course, to justify statements corresponding to Lemmas 1 and 4 without relying on the solvability of the model is a formidably difficult task.
We nevertheless stress that the irreversibility observed in the free fermion chain should be continuously connected to thermalization in non-integrable models.
It is plausible that the time evolution of the coarse-grained density distribution is insensitive to a small perturbation that breaks the integrability, such as the nearest-neighbor interaction.
It is challenging to first try establishing such stability and then proceed, in the future, to show that the perturbed model exhibits much more robust thermalization.
It would be exciting if the recent proof of the absence of local conserved quantities in certain quantum models \cite{Shiraishi2019,Shiraishi2024}, which is among the few rigorous results that apply exclusively to non-integrable systems, provides a hint for such studies.

The present theory of irreversible behavior does not provide any information about the time scale, namely, ``sufficiently large'' $T$ that appears in Theorem~3.
Although it is of essential importance to control the time scale, we still have no general results in this direction.
However, if the time-dependent state is written in the Slater determinant form
$\ket{\Phi(t)}=\{\prod_{j=1}^d\hat{d}^\dagger_j(t)\}\vac$, where $\hat{d}^\dagger_j(t)$ are the creation operators for mutually-orthogonal single-particle time-dependent states, then we may repeat the argument in the proof of Lemma~4 to get a certain estimate on the time-dependence of $\bra{\Phi(t)}\Pneq\ket{\Phi(t)}$.
See \cite{GluzaEisertFarrelly2019} for a related result on the time scale of equilibration in a free fermion system.

Finally, let us make a brief comment about the emergence of time's arrow \cite{Lebwitz1993}.
Suppose that one chooses the initial state $\kPz$ in which all particles are confined, say, in the interval $I_1$, i.e., $\hrho_1\kPz=m\rho_0\kPz$ and $\hrho_j\kPz=0$ for $j\ge2$.
Theorem 4 states that, for any $t\in[0,T]\backslash\calA$,  the measurement result of $\hrho_j$ in state $\ket{\Phi(t)}$ almost equals $\rho_0$ for every $j$.
This clearly shows that there is a directed transition from a localized density distribution at $t=0$ to a uniform distribution at large and typical $t$.
To see the implication of time-reversal symmetry, pick $T_0\in[0,T]\backslash\calA$ and set a new initial state as $\ket{\widetilde{\Phi}(0)}=\ket{\Phi(T_0)}^*$, where $*$ denotes the combination of complex conjugation and reflection $x\to L+1-x$.
Then, time-reversal invariance implies that the particles are localized in $I_1$ in the time-evolved state $\ket{\widetilde{\Phi}(T_0)}=e^{-i\hH T_0}\ket{\widetilde{\Phi}(0)}$, which is nothing but $\kPz^*$.
This seems like a forbidden transition from a uniform density distribution to a localized distribution.
Of course, there is no contradiction.
Theorem~4 shows that, for any sufficiently large $\tilde{T}$, there is an atypical set $\widetilde{\calA}\subset[0,\tilde{T}]$ for $\ket{\widetilde{\Phi}(0)}$, and $\ket{\widetilde{\Phi}(t)}=e^{-i\hH t}\ket{\widetilde{\Phi}(0)}$ have almost uniform density distribution at any $t\in[0,\tilde{T}]\backslash\widetilde{\calA}$.
Clearly $T_0$ does not belong to $[0,\tilde{T}]\backslash\widetilde{\calA}$.
We see that the time $T_0$ is atypical with respect to the initial state $\ket{\widetilde{\Phi}(0)}$.
This resolves the ``paradox''.
Another resolution, which we do not adopt here, is to introduce a probability distribution to the initial state and focus on properties that are valid for typical initial states as in classical systems \cite{Lebwitz1993,BievreParris2017}.

\medskip
{\small
It is a pleasure to thank Shelly Goldstein, Joel Lebowitz, Marcos Rigol, and Naoto Shiraishi for their stimulating discussions.
The present research is supported by JSPS Grants-in-Aid for Scientific Research No. 22K03474.
}



\begin{thebibliography}{10}

\bibitem{Lebwitz1993}
J.L. Lebowitz, {\em Boltzmann entropy and time's arrow}\/,
Physics Today, September 1993, 32--38 (1993).


\bibitem{BievreParris2017}
S. De Bievre and P.E. Parris,
{\em A rigorous demonstration of the validity of Boltzmann's scenario for the spatial homogenization of a freely expanding gas and the equilibration of the Kac ring}\/,
J. Stat. Phys. {\bf 168}, 772--793 (2017).
\\\url{https://arxiv.org/abs/1701.00116}


\bibitem{RigolMuramatsuOlshanii2006}
M. Rigol, A. Muramatsu, and M. Olshanii,
{\em Hard-core bosons on optical superlattices: Dynamics and relaxation in the superfluid and insulating regimes}\/,
Phys. Rev. A {\bf 74}, 053616 (2006).
\\\url{https://arxiv.org/abs/cond-mat/0612415}

\bibitem{RigolFitzpatrick2011}
M. Rigol and M. Fitzpatrick,
{\em Initial state dependence of the quench dynamics in integrable quantum systems}\/,
Phys. Rev. A {\bf 84}, 033640 (2011).
\\\url{https://arxiv.org/abs/1107.5811}

\bibitem{Pandeyetal}
S. Pandey, J.M. Bhat, A. Dhar, S. Goldstein, D.A. Huse, M. Kulkarni, A. Kundu, and J.L. Lebowitz,
{\em Boltzmann entropy of a freely expanding quantum ideal gas}\/,
J. Stat. Phys. {\bf 190}, article number 142, (2023).
\\\url{https://arxiv.org/abs/2303.12330}



\bibitem{GluzaEisertFarrelly2019}
M. Gluza, J. Eisert, and T. Farrelly,
{\em Equilibration towards generalized Gibbs ensembles in non-interacting theories}\/,
SciPost Phys. {\bf 7}, 038 (2019).
\\\url{https://www.scipost.org/10.21468/SciPostPhys.7.3.038}

\bibitem{ShiraishiTasaki2023}
N. Shiraishi and H. Tasaki,
{\em Nature abhors a vacuum: A simple rigorous example of thermalization in an isolated macroscopic quantum system}\/,
 J. Stat. Phys. {\bf 191}, 82 (2024).
\\\url{https://arxiv.org/abs/2310.18880}

\bibitem{LydzbaZhangRigolVidmar2021}
P.  \L yd\.{z}ba, Y. Zhang, M. Rigol, and L. Vidmar,
{\em Single-particle eigenstate thermalization in quantum-chaotic quadratic Hamiltonians}\/,
Phys. Rev. {\bf B} 104, 214203 (2021).
\\\url{https://arxiv.org/abs/2109.06895}

\bibitem{LydzbaMierzejewskiRigolVidmar2023}
P.  \L yd\.{z}ba,  M. Mierzejewski, M. Rigol, and L. Vidmar,
{\em Generalized thermalization in quantum-chaotic quadratic Hamiltonians}\/,
Phys. Rev. Lett. 131, 060401 (2023).
\\\url{https://arxiv.org/abs/2210.00016}



\bibitem{GuckenheimerHolmes}
J. Guckenheimer and Philip Holmes,
{\em Nonlinear Oscillations, Dynamical Systems, and Bifurcations of Vector Fields}\/,
(Springer, Applied Mathematical Sciences, 1983).

\bibitem{Gutzwiller}
M.C. Gutzwiller,
{\em Chaos in Classical and Quantum Mechanics}\/,
(Springer, Interdisciplinary Applied Mathematics, 1990).

\bibitem{EckmannRuelle}
J.-P. Eckmann and D. Ruelle,
{\em Ergodic theory of chaos and strange attractors}\/,
Rev. Mod. Phys. {\bf 57}, 671--656 (1985).



\bibitem{vonNeumann}
J. von Neumann,
{\em Beweis des Ergodensatzes und des $H$-Theorems in der neuen Mechanik}\/,
Z. Phys. \textbf{57}, 30 (1929);\\
English translation (by R. Tumulka),
{\em Proof of the Ergodic Theorem and the H-Theorem in Quantum Mechanics}\/, The European Phys. J.  H {\bf 35} 201--237 (2010).\\
\url{https://arxiv.org/abs/1003.2133}

\bibitem{Deutsch1991}
J.M. Deutsch,
{\em Quantum statistical mechanics in a closed system}\/,
Phys. Rev. A {\bf 43}, 2046 (1991).

\bibitem{Srednicki1994}
M. Srednicki,
{\em Chaos and quantum thermalization}\/,
Phys. Rev. E {\bf 50}, 888 (1994).

\bibitem{Tasaki1998}
H. Tasaki,
{\em 	From Quantum Dynamics to the Canonical Distribution: General Picture and a Rigorous Example}\/, 
Phys. Rev. Lett.  \textbf{80}, 1373--1376 (1998).\\
\url{https://arxiv.org/abs/cond-mat/9707253}

\bibitem{GLTZ}
S. Goldstein, J. L. Lebowitz, R. Tumulka, N. Zangh\`\i,
{\em Long-time behavior of macroscopic quantum systems: Commentary accompanying the English translation of John von Neumann's 1929 article on the quantum ergodic theorem}\/,
European Phys. J. H {\bf 35}, 173--200 (2010).\\
\url{https://arxiv.org/abs/1003.2129}

\bibitem{GLMTZ09b} 
S. Goldstein, J. L. Lebowitz, C. Mastrodonato, R. Tumulka, and N. Zangh\`\i,
{\em On the Approach to Thermal Equilibrium of Macroscopic Quantum Systems}\/,
Phys. Rev. E \textbf{81}, 011109 (2010).\\
\url{https://arxiv.org/abs/0911.1724}

\bibitem{RigolSrednicki2012}
M. Rigol and M. Srednicki,
{\em Alternatives to Eigenstate Thermalization}\/,
Phys. Rev. Lett. {\bf 108}, 110601 (2012).
\\\url{https://arxiv.org/abs/1108.0928}

\bibitem{Reimann3}
P. Reimann,
{\em Generalization of von Neumann's Approach to Thermalization}\/,
Phys. Rev. Lett. {\bf 115}, 010403 (2015).\\
\url{https://arxiv.org/abs/1507.00262}

\bibitem{DAlessioKafriPolkovnikovRigol2016}
L. D'Alessio, Y. Kafri, A. Polkovnikov, and M. Rigol,
{\em From quantum chaos and eigenstate thermalization to statistical mechanics and thermodynamics}\/,
Adv. Phys. {\bf 65},  239--362 (2016).
\\\url{https://arxiv.org/abs/1509.06411}

\bibitem{Tasaki2016}
H. Tasaki,
{\em Typicality of thermal equilibrium and thermalization in isolated macroscopic quantum systems}\/,
J. Stat. Phys. {\bf 163}, 937--997 (2016).
\\\url{https://arxiv.org/abs/1507.06479}

\bibitem{Tasaki2010}
H. Tasaki,
{\em The approach to thermal equilibrium and ``thermodynamic normality" --- An observation based on the works by Goldstein, Lebowitz, Mastrodonato, Tumulka, and Zanghi in 2009, and by von Neumann in 1929}\/, (unpublished note 2010).
\\\url{https://arxiv.org/abs/1003.5424}

\bibitem{RoosTeufelTumulkaVogel2024}
B. Roos, S. Teufel, R. Tumulka, and C. Vogel,
{\em Macroscopic Thermalization for Highly Degenerate Hamiltonians}\/,
(preprint, 2024).\\
\url{https://arxiv.org/abs/2408.15832}

\bibitem{Shiraishi2019}
N. Shiraishi, 
{\em Proof of the absence of local conserved quantities in the XYZ chain with a magnetic field}\/, 
Europhys. Lett. {\bf 128} 17002 (2019).
\\\url{https://arxiv.org/abs/1803.02637}

\bibitem{Shiraishi2024}
N. Shiraishi, 
{\em Absence of Local Conserved Quantity in the Heisenberg Model with Next-Nearest-Neighbor Interaction}\/,
J. Stat. Phys. {\bf 191}:114 (2024).
\\\url{https://link.springer.com/article/10.1007/s10955-024-03326-4}


\end{thebibliography}
\end{document}